\documentclass[aps,prd,amsmath,amssymb,amsfonts,floats,floatfix,superscriptaddress,nofootinbib,notitlepage]{revtex4-1}

\usepackage{graphicx}
\usepackage{graphics}
\usepackage[retainorgcmds]{IEEEtrantools}
\usepackage{color}
\usepackage{xspace} 
\usepackage{slashed}
\usepackage[colorlinks]{hyperref}
\usepackage{comment}
\usepackage{graphicx}

\newcommand{\leaveout}[1]{}

\newcommand{\sMtwo}{\sigma_2}

\newcommand{\CS}{\mathcal{M}_{\times}}

\newcommand{\indmode}{\ell}

\newcommand{\VVc}{\hat\Delta}
\newcommand{\VVtd}{\Delta_{2d}}
\newcommand{\VVsph}{\Delta_{\mathbb{S}^2}}

\newcommand{\sco}{\hat\sigma}

\newcommand{\Mtwo}{\mathcal{M}_2}
\newcommand{\mt}{\mathcal{M}_\times}

\newcommand{\Gret}{G_{\text{ret}}}
\newcommand{\Glret}{G^\text{ret}_{\indmode}}
\newcommand{\Gld}{G^\text{d}_{\indmode}}
\newcommand{\Gnd}{G_{\text{nd}}}
\newcommand{\Gd}{G_{\text{d}}}

\newcommand{\coord}[1]{#1}
\newcommand{\dt}{\Delta t}
\newcommand{\dr}{\Delta r}
\newcommand{\dx}{\Delta x}

\newcommand{\nn}{{\mathcal{N}}}

\begin{document}
\global\parskip 6pt

\author{Marc Casals}
\email{mcasals@cbpf.br, marc.casals@ucd.ie}
\affiliation{Centro Brasileiro de Pesquisas F\'isicas (CBPF), Rio de Janeiro, 
CEP 22290-180, 
Brazil.}
\affiliation{School of Mathematics and Statistics, University College Dublin, Belfield, Dublin 4, Ireland.}

\author{Brien C. Nolan}
\email{brien.nolan@dcu.ie}
\affiliation{School of Mathematical Sciences, Dublin City
University, Glasnevin, Dublin 9, Ireland.}

\author{Adrian C. Ottewill}
\email{adrian.ottewill@ucd.ie}
\affiliation{School of Mathematics and Statistics, University College Dublin, Belfield, Dublin 4, Ireland.}

\author{Barry Wardell}
\email{barry.wardell@ucd.ie}
\affiliation{School of Mathematics and Statistics, University College Dublin, Belfield, Dublin 4, Ireland.}

\title{Regularized calculation of the retarded Green function in Schwarzschild spacetime}

\begin{abstract} 
The retarded Green function for linear field perturbations of black hole spacetimes is notoriously
difficult to calculate. One of the difficulties is due to a Dirac-$\delta$ divergence that the
Green function possesses when the two spacetime points are connected by a ``direct" null geodesic.
We present a procedure which notably aids its calculation in the case of Schwarzschild spacetime by separating this direct
$\delta$-divergence from the remainder of the retarded Green function. More precisely, the method 
consists of calculating the multipolar $\ell$-modes of the direct $\delta$-divergence and
subtracting them from the corresponding modes of the retarded Green function. We illustrate the
usefulness of the method with some specific calculations in the case of the scalar Green function
and self-field for a point scalar charge in Schwarzschild spacetime.
\end{abstract}

\date{\today}
\maketitle


\section{Introduction}

Linear field perturbations of black hole spacetimes obey a wave equation. The retarded Green
function (GF) of this equation is an important function of two spacetime points, as it serves to
evolve any initial field data to its future. Certain problems require knowing the retarded Green
function {\it globally}, i.e., for points arbitrarily separated. For example, the self-field on a
point particle moving on a background black hole spacetime can be expressed in terms of an integral
of the GF over the past worldline of the particle. The self-force can then be obtained as a
derivative of the self-field (see \cite{Poisson:2011nh,Wardell:2015kea} for reviews). Also, within
the different setting of relativistic quantum information, the probability of a particle detector
being excited by a field emitted by another detector moving on a curved background can be expressed
as a (double) integral of the GF\footnote{Even though in this setting of relativistic quantum
information the field is quantized, to {\it leading} order in the coupling between the detectors
and the field, the signal strength depends only on the {\it retarded} Green function and so it does
not depend on the quantum state of the field.} (see,
e.g.,~\cite{jonsson2015information,blasco2015violation}).

Calculating the GF on a black hole background is no easy endeavour. One of the difficulties lies in
the fact that the GF diverges when the two spacetime points are connected via a null
geodesic~\cite{Garabedian,Ikawa}. The so-called Hadamard form shows that, for the case of ``direct"
null geodesics (i.e., for null geodesics which have not orbited around the black hole and so, in
particular, have not encountered a caustic), this divergence is of a Dirac-$\delta$
type~\cite{Hadamard,Friedlander,Poisson:2011nh}. The term in the Hadamard form which contains this
direct divergence is called the {\it direct part}. In a practical calculation, where the GF is only
calculated {\it approximately} to within a desired accuracy, this direct divergence is
typically smeared out and ``contaminates" the evaluation of the GF even when the spacetime points
are {\it timelike}-separated.

In this paper, we present a simple but very useful idea for facilitating the practical evaluation
of the GF on a static and spherically-symmetric spacetime (including, for example, the
Schwarzschild black hole spacetime) via a multipolar $\ell$-mode decomposition (as used, for
examle, in Refs.~\cite{CDOW13,PhysRevD.89.084021}). Within such a decomposition, the
divergences of the GF when the spacetime points are null-separated manifest as divergences of the
infinite sum over $\ell$-modes~\cite{casals2016global}. Since, in a practical calculation, one must
truncate the sum at a {\it finite} number of modes, the divergences of the computed GF are
inevitably spread out. This implies, in particular, that when the points are close to being
connected by a direct null geodesic, a large number of $\ell$-modes are required in the sum in
order to avoid contamination from the direct part. Our proposal is to obtain the $\ell$-modes
of the direct part in the Hadamard form and subtract them from the $\ell$-modes of the full GF,
prior to carrying out the $\ell$-sum. The resulting object is, thus, essentially, the full GF {\it
minus} the direct part.  Since this object does not diverge when the points are connected by a
direct null geodesic, its $\ell$-sum converges much faster for points close to being connected by
such a geodesic.

The contribution from the direct part is, in fact, not needed for certain problems such as the
self-force problem (in this case, the direct part is ``regularized away"). For problems where this
contribution is needed, such as in the relativistic quantum information setting, it can be
calculated separately using an alternative method which does not involve an $\ell$-mode
decomposition. Although we present the method explicitly for the case of a massless scalar field on
Schwarzschild spacetime, it can readily be extended to fields of non-zero spin. We illustrate the
usefulness of our proposal with a practical calculation of the scalar GF and self-field in
Schwarzschild spacetime. Our example shows that many fewer $\ell$-modes are required to achieve a
certain accuracy when our proposal of subtracting the $\ell$-modes of the direct part is used, thus
greatly facilitating the evaluation of the $\ell$-sum.

This paper is organized as follows. In Sec.~\ref{sec:GF} we present the GF for the wave equation in
curved spacetimes and the difficulties with its practical evaluation in Schwarzschild spacetime. In
Sec.~\ref{sec:Proposal} we present our proposal for facilitating the evaluation of the GF. We
implement this proposal specifically in the case of the calculation of the scalar GF and self-field
in Sec.~\ref{sec:application}. We finish in Sec.~\ref{sec:extension} with possible extensions of
the application of our method. In the appendices we derive small-coordinate expansions of the
$\ell$-modes of the retarded Green function and of its direct part.

We use geometric units $c=G=1$ and metric signature $(- + + +)$ throughout this work.


\section{Green function}\label{sec:GF}

Let us consider a massless scalar field propagating on a curved background spacetime, with the
field satisfying the Klein-Gordon equation. The corresponding retarded Green function (GF)
satisfies the Klein-Gordon equation with a four-dimensional (invariant) Dirac-$\delta$ distribution as the
source:
\begin{equation}
\Box_{\coord{x}} \Gret(\coord{x},\coord{x}')=-4\pi \frac{\delta_4(\coord{x}-\coord{x}')}{\sqrt{-g(\coord{x})}},
\end{equation}
where $\coord{x}'$ and $\coord{x}$ are two spacetime points\footnote{As is common, we blur the
distinction between points (e.g. $x$ and $x'$) and their coordinates given a global coordinate
system on the exterior Schwarzschild spacetime under consideration.} and $\Box_{\coord{x}}$ is the
D'Alembertian with respect to $\coord{x}$. The GF obeys causal boundary conditions: it is equal to
zero if $\coord{x}'$ does not lie in the causal past of $\coord{x}$.

There exists an analytical expression for the GF, the so-called {\it Hadamard form}, which is valid
when $\coord{x}'$ is in a local neighbourhood of $\coord{x}$. More precisely, $\coord{x}'$ must lie
in a {\it normal neighbourhood} $\mathcal{N}(x)$ of $\coord{x}$: a region $\mathcal{N}(x)$
containing $\coord{x}$ such that every $\coord{x}'\in \mathcal{N}(x)$ is connected to $\coord{x}$
by a {\it unique} geodesic which lies in $\mathcal{N}(x)$. The Hadamard form
is~\cite{Hadamard,Friedlander,Poisson:2011nh}:
\begin{equation}\label{eq:hadamard}
G_{ret}(\coord{x},\coord{x}') =\big[U(\coord{x},\coord{x}')\delta(\sigma)+ V(\coord{x},\coord{x}')\theta(-\sigma)\big]
\theta(\dt),
\end{equation} 
where $U(x,x')$ and $V(x,x')$ are regular, real-valued biscalars, $\dt\equiv t-t'$ and $t$ is a
time coordinate. Here, $\sigma=\sigma(\coord{x},\coord{x'})$ is Synge's
world-function~\cite{Synge}, which is equal to one-half of the squared distance along the (unique)
geodesic connecting $\coord{x}$ and $\coord{x}'$. This means that $\sigma$ is
negative/zero/positive whenever that geodesic is timelike/null/spacelike. Eq.~\eqref{eq:hadamard}
exhibits a Dirac-$\delta$ divergence at $\sigma=0$, i.e., when the spacetime points are connected
by a ``direct" null geodesic or else when they coincide (i.e., $\coord{x'}=\coord{x}$). We thus
refer to $\Gd(\coord{x},\coord{x}')\equiv U\delta(\sigma)\theta(\dt)$, which only has support when
the points are null separated, as the ``direct part", and to $V\theta(-\sigma)\theta(\dt)$, which also
has support when the points are timelike separated, as the ``tail part". The ``direct biscalar" $U(x,x')$ is related to the
so-called van Vleck determinant $\Delta$~\cite{VanVleck:1928,Morette:1951,Visser:1993} by
$U(\coord{x},\coord{x'})=\Delta^{1/2}(\coord{x},\coord{x'})$ and satisfies a transport equation
along the unique geodesic joining $\coord{x}$ and $\coord{x}'$. In turn, the ``tail biscalar"
$V(x,x')$ satisfies the homogeneous Klein-Gordon equation.

As mentioned, the Hadamard form is only valid in a local neighbourhood, while one may need the GF
for arbitrarily separated points in the spacetime. When the background spacetime is
spherically-symmetric, it is useful to carry out a multipolar decomposition of the GF, which is
valid {\it globally}, as 
\begin{align} \label{eq:Green}
&
\Gret(\coord{x},\coord{x'})=
\frac{1}{r\, r'}
\sum_{\ell=0}^{\infty}(2\ell+1)P_{\ell}(\cos\gamma)\Glret(r,r';\dt),
\end{align}
where $\gamma\in[0,\pi]$ is the angle separation between $\coord{x}$ and $\coord{x}'$ and $r$ is a
radial coordinate. The multipolar modes $\Glret$ satisfy a Green function equation in
$(1+1)$-dimensions. These modes can be calculated in a variety of ways -- for example, as a Fourier
integral over real frequencies~\cite{GFKerr}; as a sum over quasi-normal modes plus a branch cut
integral~\cite{Leaver:1986,CDOW13}; or via a numerical integration of the $(1+1)$-D Green function
equation that they satisfy~\cite{PhysRevD.89.084021,Mark:2017dnq}.

When calculating the GF via Eq.~\eqref{eq:Green}, the Dirac-$\delta$ divergence at $\sigma=0$
arises from a divergence in the infinite $\ell$-mode sum~\cite{casals2016global}. In a practical
calculation, however, it is not possible to include an infinite number of $\ell$-modes and so the
infinite sum must be truncated at some finite upper cutoff. As a consequence of this finite cutoff,
spurious oscillations appear, but these can be smoothed out via the introduction of a factor which
decays fast for large $\ell$~\cite{CDOW13,Hardy}. The finite cutoff, including a smoothing
factor, effectively means that the sharp $\delta(\sigma)$ divergence is not exactly captured and,
instead, one obtains an element of a $\delta$-convergent sequence. For example, in this paper, we
chose $e^{-\ell^2/(2\ell_{cut}^2)}$, for some choice of $\ell_{cut}\in \mathbb{R}$, as the
smoothing factor, and then the $\delta$-convergent sequence is a Gaussian distribution centered at
$\sigma=0$; we shall henceforth refer to the $\delta$-convergent sequence as a Gaussian
distribution for simplicity. This means that the approximate GF resulting from the finite cutoff
and smoothing factor is highly ``contaminated" by the Gaussian distribution at points near
$\sigma=0$. Therefore, in a normal neighbourhood $x'\in \mathcal{N}(x)$, even though the {\it
exact} $G_{ret}(\coord{x},\coord{x}')$ is equal to $V(x,x')$ for points $\coord{x}'$
timelike-separated from $\coord{x}$, the {\it approximate} $G_{ret}(\coord{x},\coord{x}')$ may
differ considerably from the correct value given by $V(x,x')$ for points near $\sigma=0$. This
approximation to the GF becomes worse the closer the points are to $\sigma=0$.

In previous work \cite{Poisson:Wiseman:1998,Anderson:Wiseman:2005,CDOWa,CDOW13,PhysRevD.89.084021},
the above problem of obtaining the GF for points ``near" $\sigma=0$ was addressed by calculating
the GF in that regime, not via Eq.~\eqref{eq:Green}, but rather via a direct evaluation of
$V(x,x')$. Such evaluation can be achieved, for example, using the following multiple power
series~\cite{Ottewill:Wardell:2008}:
\begin{equation}\label{eq:V power series}
V(\coord{x},\coord{x}')=\sum_{i,j,k=0}^{\infty}v_{ijk}(r)
(t-t')
^{2i}(1-\cos\gamma)^j(r-r')^k,
\end{equation}
for some coefficients $v_{ijk}$ that can be determined. Again, in a practical calculation, one must
stop the sums in \eqref{eq:V power series} at some finite upper limits, thus yielding an
approximation to the regular bitensor $V(x,x')$. The approximation becomes worse the further the
points are from each other in the sense of large coordinate increments (i.e.\ large $|t-t'|$,
$|r-r'|$ and/or $\gamma$) \footnote{We note that such points where the truncated approximation to
Eq.~\eqref{eq:V power series} becomes worse may still be close to $\sigma=0$, as well as including
points for which $\sigma(\coord{x},\coord{x}')$ is large.}. The aim is to match the calculation of
$V(x,x')$ via \eqref{eq:V power series}, with the sums truncated, to the mode-sum calculation via
\eqref{eq:Green}, also with the sum truncated, in a region where both approximations are accurate
enough (i.e., the truncation error is smaller than a desired accuracy). The region where the
approximation to $V(x,x')$ is accurate enough is called the quasi-local (QL) region, and the region
where the approximation to the mode sum \eqref{eq:Green} is accurate enough is called the Distant
Past (DP).

This approach encounters a fundamental challenge, namely that the existence of an overlap between the
QL and the DP regions is not guaranteed. In Ref.~\cite{CDOW13,PhysRevD.89.084021} it was shown
that, by using a Pad\'e resummation of the QL approximation, the two regions do overlap for many
cases of relevance to the self-force problem in Schwarzschild spacetime. However, the overlapping
region was found to be often quite small, and ultimately the matching between QL and DP regions was
found to be the dominant limitation in achieving robust (i.e. valid in a wide variety of scenarios)
and accurate results. In order to increase the chances of having a region of overlap, or to improve
agreement in an overlapping region, one has two options: (i) to improve the calculation of
$V(x,x')$ so that the QL approximation becomes more accurate in a larger region; (ii) to improve
the calculation via \eqref{eq:Green} so that the DP approximation becomes more accurate in a larger
region. The former has a fundamental limitation, which is that $V(x,x')$ diverges at the edge of
the normal neighbourhood, and so any power series approximation will struggle to accurately
represent it near this divergence. The latter saw some improvement by going from a quasinormal mode
sum approximation \cite{CDOW13}, which is known to have very poor convergence when the points are
close together, to a time domain method for computing $\Glret$ \cite{PhysRevD.89.084021}. However,
this improvement did not address the problem of poor convergence of the $\ell$-sum in the DP
contribution. In this paper we present a method for addressing this problem, largely eliminating it.


\section{Regularized calculation of the Green function} \label{sec:Proposal}

As mentioned, the calculation of the GF via \eqref{eq:Green} encounters problems for points near 
$\sigma=0$ because the truncation of the sum (in combination with the introduction of a 
smoothing factor) effectively turns the Dirac-$\delta$ divergence exhibited in \eqref{eq:hadamard} 
into a widespread Gaussian distribution.
One might expect that the approximation will improve if we
calculate a multipolar decomposition of the result of subtracting the direct part $\Gd(x,x')\equiv
U\delta(\sigma)\theta(\dt)$ from the GF. The reason is that the quantity resulting from such
subtraction (which is equal to $V\theta(-\sigma)\theta(\dt)$ when $x'\in \mathcal{N}(x)$) is finite
at $\sigma=0$. In order to achieve this, in this section we shall subtract the multipolar modes of
the direct part $\Gd$ from the multipolar modes $\Glret$ of the full GF. For this purpose, it is
very convenient to follow Ref.~\cite{casals2016global}.
 
From now on we focus on Schwarzschild space-time with mass $M$, and will work in Schwarzschild
coordinates $\{t,r,\theta,\varphi\}$. First, we make the conformal transformation
\begin{equation}
  d\hat{s}^2 \equiv r^{-2}ds^2 =
   ds_2^2g
   +d\Omega_2^2,
  \label{con-sch-lel}
\end{equation}
where $ds^2$ and $d\Omega_2^2$ are the line-elements in, respectively, Schwarzschild spacetime and
the unit 2-sphere $\mathbb{S}^2$, and we have defined
\begin{equation}
  ds_2^2 \equiv 
  \frac{f}{r^2}\left(-dt^2+f^{-2}dr^2\right),
  \label{2d-lel}
\end{equation}
with $f\equiv 1-2M/r$. We call the 4-D spacetime with the metric \eqref{con-sch-lel} the conformal
Schwarzschild spacetime $\CS$, and we call the spacetime with metric \eqref{2d-lel} the 2-D
conformal spacetime $\Mtwo$. This 2-D spacetime was studied in~\cite{PhysRevD.92.104030}, where it
was proven to be a causal domain (i.e., $\Mtwo$ is geodesically convex and it obeys a certain
causality condition)~\cite{Friedlander}. As a consequence, both the world-function and the Hadamard
form for the Green function in $\Mtwo$ are valid globally in this 2-D spacetime~\cite{Friedlander}.

Since Eq.~\eqref{con-sch-lel} is a conformal transformation, if we consider a conformally-coupled
scalar field we have that the GF in Schwarzschild spacetime is~\cite{Friedlander,Birrell:Davies}
\begin{equation} 
  \Gret(\coord{x},\coord{x'}) = \frac{1}{r\cdot r'}\hat{G}_R(\coord{x},\coord{x'}),
  \label{con-gf}
\end{equation}
where $\hat{G}_R(\coord{x},\coord{x'})$ is the GF for the Klein-Gordon equation in conformal
Schwarzschild spacetime. Both GF's $\Gret$ and $\hat{G}_R$ admit the Hadamard form
\eqref{eq:hadamard}; we denote the biscalars in Schwarzshild spacetime as in \eqref{eq:hadamard},
whereas we denote the biscalars in the conformal Schwarzshild spacetime with the corresponding
hatted symbols, so that, in particular, $\VVc(\coord{x},\coord{x'})$ and $\sco$ denote,
respectively, the van Vleck determinant and the world function in $\CS$. By comparing the direct
parts of the Hadamard forms for the two GF's, we have that
\begin{equation}\label{eq:con-d}
  \Gd(\coord{x},\coord{x'})= \Delta^{1/2}(\coord{x},\coord{x'})\delta(\sigma)\theta(\dt)= \frac{\VVc^{1/2}(\coord{x},\coord{x'})\delta(\sco)\theta(\dt)}{r\cdot r'}.
\end{equation}
Equation \eqref{eq:con-d} holds because $\sigma(\coord{x},\coord{x'})=0$ if and only if
$\sco(\coord{x},\coord{x'})=0$, which follows from the invariance properties of null geodesics
under conformal transformations.

As mentioned, the Hadamard form \eqref{eq:hadamard} is only valid in normal neighbourhoods [we
will later explicitly determine the normal neighbourhood of an arbitrary point in $\CS$: the result
is given in Eq.~\eqref{eq:NN-x}]. A great advantage of the transformation in \eqref{con-sch-lel} is
that the conformal Schwarzschild spacetime is a direct product: the manifold has the form
${\CS}=\Mtwo\times \mathbb{S}^2$, with line element given by (\ref{con-sch-lel}) with
(\ref{2d-lel}). Writing coordinates on the 4-D spacetime ${\CS}$ as $x^\alpha =(x^A,x^a)$
where $x^A, A=0,1$ are coordinates on $\Mtwo$ and $x^a, a=2,3$ are coordinates on the 2-sphere
$\mathbb{S}^2$, we see that the metric tensor decomposes as \footnote{Greek letters as indices
denote indices in the 4-D spacetime, capital Latin letters indices on $\Mtwo$ and small Latin
letters indices on $\mathbb{S}^2$.}
\begin{equation}
  g_{\alpha\beta} = \left(\begin{array}{cc} g_{AB}(x^C) & 0 \\ 0 & g_{ab}(x^c) \end{array} \right), \label{product-metric}
\end{equation}
while the (Levi-Civita) connection coefficients decompose as 
\begin{equation}
  {\Gamma^A}_{\beta\gamma}(x^\delta) = \left(\begin{array}{cc} {\Gamma^A}_{BC}(x^D) & 0 \\ 0 & 0 \end{array} \right), \quad
  {\Gamma^a}_{\beta\gamma}(x^\delta) = \left(\begin{array}{cc} 0 & 0 \\ 0 & {\Gamma^a}_{bc}(x^d) \end{array} \right). 
  \label{product-connection} 
\end{equation}
It then follows that there is a one-to-one correspondence between geodesics of the spacetime
${\CS}$ and `products' of geodesics on $\Mtwo$ and $\mathbb{S}^2$. That is, a parametrized curve
$\lambda:x^\alpha=x^\alpha(s) = (x^A(s),x^a(s))$ is a geodesic of the spacetime ${\CS}$ if and only
if $\mu:x^A=x^A(s)$ and $\nu:x^a=x^a(s)$ are geodesics on $\Mtwo$ and $\mathbb{S}^2$ respectively.
(We will say that $\mu$ and $\nu$ \textit{lift} to yield $\lambda$, which \textit{projects} to
yield $\mu$ and $\nu$.) As noted above, $(\Mtwo,g_{AB})$ is geodesically convex: there is a unique
geodesic between any pair of points of $\Mtwo$. Then the world function $\sMtwo$ on $\Mtwo$ is defined
globally and may be written as
\begin{equation}
  \sMtwo(x^A,x^{A'}) = \frac{\epsilon}{2}\eta^2, \label{sigma1}
\end{equation}
where $\epsilon = -1, 0, +1$ for timelike, null and spacelike separations in $\Mtwo$ and
$\eta$ is proper time, zero and proper distance respectively along the corresponding timelike, null
and spacelike geodesics. (By convention, $\eta\geq0$ along a future-directed causal curve from
$x^{A'}$ to $x^{A}$.) Below Eq.~\eqref{eq:Green}, we already defined
$\gamma=\gamma(x^a,x^{a'})\in[0,\pi]$, which we can here understand more explicitly as the
\textit{geodesic separation} of $x^a$ and $x^{a'}$: the proper length of the \textit{shortest}
geodesic from $x^a$ to $x^{a'}$.
We distinguish this from the \textit{geodesic distance} ${\gamma}_\nu(x^a,x^{a'})$ which measures
the length of a geodesic $\nu$ from $x^a$ to $x^{a'}$. This may be arbitrarily long (by running
around the sphere multiple times) and may also run in the direction opposite to that of the
shortest geodesic. Note that we must have either
\begin{equation}
  {\gamma}_\nu(x^a,x^{a'})=\gamma+2n\pi
\end{equation}
or
\begin{equation}\label{eq:eta=2npi-gamma}
  {\gamma}_\nu(x^a,x^{a'})=2n\pi-\gamma
\end{equation}
for some $n\in\mathbb{N}$.

The question of uniqueness of geodesics connecting points of the 4-D spacetime is then easily
resolved \cite{casals2016global}. Consider two points $x=(x^A,x^a)$ and $x'=(x^{A'},x^{a'})$ of the
spacetime ${\CS}$, where $x^a$ and $x^{a'}$ are non-antipodal points on $\mathbb{S}^2$. There is a unique geodesic $\mu$ of $\Mtwo$ connecting $x^A$ and $x^{A'}$, and
generically 
there are two countably infinite families of
geodesics on $\mathbb{S}^2$ connecting $x^a$ and $x^{a'}$. Among these there is a unique shortest
geodesic $\nu_0$ for which ${\gamma}_{\nu_0}(x^a,x^{a'})=\gamma(x^a,x^{a'})$. In the case of a null
geodesic, the pair $\mu$, $\nu_0$ lift to what we have referred to as the \textit{direct} null
geodesic connecting points of
Schwarzschild spacetime (in fact the conformal transformation from ${\mathcal{M}}_\times$ to
Schwarzschild is also required). These geodesics on $\Mtwo$ and $\mathbb{S}^2$ lift to give
geodesics on the spacetime ${\CS}$ as noted above, and so spacetime points are (generically)
connected by two countably infinite families of geodesics. Among these, the geodesic formed by
lifting $\mu$ and $\nu_0$ is distinguished. We can then define a 2-point function on the spacetime
${\mathcal{M}}_\times$ by
\begin{equation}
  \sco(\coord{x},\coord{x'}) = 
      \frac{\epsilon}{2}\eta^2
    + \frac12\gamma^2.\label{world-con}
\end{equation}
Care is needed in interpreting this as the world function of the spacetime as $\sco$ measures one
half the square of the geodesic distance (appropriately signed) along the unique geodesic whose
projection into $\mathbb{S}^2$ yields $\nu_0$. This is not the only geodesic from $x$ to $x'$.
However, this situation changes when we restrict to \textit{causal} geodesics. So consider a causal
geodesic $\lambda:x^\alpha=x^\alpha(s)$ from $x$ to $x'$ in ${\CS}$. This projects to a causal
geodesic $\mu:x^A=x^A(s)$ on $\Mtwo$ and a spacelike geodesic $\nu:x^a=x^a(s)$ on $\mathbb{S}^2$.
Then one half the square of the geodesic distance from $x$ to $x'$ along $\lambda$ is given by
\begin{equation}
  \hat{\sigma}_\lambda(x,x') = \frac{\epsilon}{2}\eta^2(x^A,x^{A'}) + \frac12\gamma_\nu^2(x^a,x^{a'}). \label{geo-dist}
\end{equation}
This must be non-positive, and so there is a \textit{finite} number of geodesics $\nu$ which allow
for $\lambda$ to be causal, among which must be the shortest spacelike geodesic $\nu_0$. This leads
us to a crucial point: Suppose that $x=(x^A,x^a), x'=(x^{A'},x^{a'}) \in{\CS}$ are connected by a
causal geodesic which is not a radial null geodesic $(\gamma(x^a,x^{a'})\neq0)$. (It is easily
verified that spacetime points connected by a radial null geodesic are connected by a unique causal
geodesic.) Then $\eta$, the proper time separation of (necessarily) timelike separated points $x^A$
and $x^{A'}$ in $\Mtwo$ is fixed, and we must have $\eta(x^A,x^{A'})\geq
\gamma(x^a,x^{a'})=\gamma_{\nu_0}$. The next shortest geodesic $\nu_1$ on ${\mathbb{S}}^2$
connecting $x^a$ and $x^{a'}$ lies on the great circle connecting these points, running in the
direction opposite to that of $\nu_0$, and so has $\gamma_{\nu_1}=2\pi-\gamma_{\nu_0}=2\pi-\gamma$.
The corresponding geodesic $\lambda_1=(\mu,\nu_1)$ is causal if and only if $\eta\geq2\pi-\gamma$.
If $\lambda_1$ is not causal, then no further spacelike geodesic $\nu_*$ of ${\mathbb{S}}^2$ from
$x^a$ to $x^{a'}$ - which would have $\gamma_{\nu_*}>\gamma_{\nu_1}$ - will yield a causal geodesic
$\lambda_*=(\mu,\nu_*)$ of ${\mathcal{M}}_\times$. Hence the inequalities $\eta\geq\gamma$ and
$\eta\geq 2\pi-\gamma$ are the necessary and sufficient conditions for the existence of multiple
causal geodesics from $x$ to $x'$. Note that the conditions that both $\lambda=(\mu,\nu_0)$ and
$\lambda_1=(\mu,\nu_1)$ are causal imply that $\eta\geq\pi$. For $\eta(x^A,x^{A'})<2\pi-\gamma$ and
given $x=(x^A,x^a)$, there is a non-trivial open set of points $x'$ for which $x$ and $x'$ are
connected by a unique causal geodesic. In
order to obtain an open set of points $x'$ connected to $x$ by a unique geodesic (regardless of
causal character) which remains within that set, we must exclude the point
$x^{a'}\in{\mathbb{S}}^2$ antipodal to $x^a$, which we did not consider in the whole argument above. That is, each $x=(x^A,x^a)\in{\mathcal{M}}_\times$
has a (maximal) normal neighbourhood $\nn(x)$ of the form
\begin{equation}
  \nn(x) = \{(x^{A'},x^{a'})\in\mt: \eta(x^A,x^{A'})<2\pi-\gamma(x^a,x^{a'}), \gamma(x^a,x^{a'})<\pi\}. \label{eq:NN-x}
\end{equation}
Then, for each $x\in\mt$, (\ref{world-con}) defines the world function $\hat{\sigma}(x,x')$ on
$\nn(x)$, the retarded Green function $\hat{G}_R(x,x')$ assumes the Hadamard form corresponding to
(\ref{eq:hadamard}) and the direct part takes the form given in (\ref{eq:con-d}).
Existence, uniqueness and the corresponding
form to Eq. \eqref{eq:hadamard} for $\hat{G}_R$ follow from Theorem 4.5.1 and Corollary
5.1.1 of \cite{Friedlander}\footnote{The relevant results require that $(x,x')$ lie in a causal
domain $\Omega$ of the spacetime, for which all pairs of points of $\Omega$ are joined by a unique
geodesic which lies in $\Omega$. We can construct the required geodesically convex regions by
excluding a semi-great circle of points from ${\mathcal{S}}^2$ rather than a single point. Since
the base point $x$ is fixed throughout, we can keep the discussion in terms of normal
neighbourhoods rather than causal domains.}.

Another advantage of conformal Schwarzschild spacetime being a direct product is that its
van Vleck determinant may be factorized as
(e.g.,~\cite{Casals:2012px})
\begin{equation}\label{eq:VVc}
  \VVc(\coord{x},\coord{x'})=\VVtd(x^A,x^{A'})\VVsph(\gamma),
\end{equation}
where $\VVtd$ and
\begin{equation}\label{eq:VVsph}
  \VVsph\equiv \frac{\gamma}{\sin\gamma}
\end{equation} 
are the van Vleck determinants in, respectively, $\Mtwo$ and $\mathbb{S}^2$.

After determining normal neighbourhoods in $\CS$ and the explicit $\gamma$-dependence of the direct part  $\Gd=U\delta(\sigma)\theta(\dt)$ in Schwarzschild spacetime, we can now proceed to calculate the multipolar
modes of this direct part. 
From
Eqs.~\eqref{eq:con-d}, \eqref{eq:VVc} and \eqref{eq:VVsph}, these multipolar modes are given by
\begin{align}\label{eq:Gld int}
  \Gld(r,r';\dt)& \equiv 
    \frac{r\, r'}{2}
  \int_{-1}^{+1} d(\cos\gamma) P_{\ell}(\cos\gamma)\Delta^{1/2}(\coord{x},\coord{x'})\delta(\sigma)\theta(\dt)
  \nonumber \\ &
  =
\frac{\theta(\dt)}{2}
  \VVtd^{1/2}(x^A,x^{A'}) \int_{-1}^{+1}d(\cos\gamma) P_{\ell}(\cos\gamma)\left(\frac{\gamma}{\sin\gamma}\right)^{1/2}\delta(\sco).
\end{align}
Using \eqref{world-con}, we have that 
\begin{equation} \label{eq:Gld}
  \Gld(r,r';\dt)=
\frac{\theta(\dt)}{2}
   \theta(\pi-\eta) \VVtd^{1/2}(x^A,x^{A'}) P_{\ell}(\cos\eta)\left(\frac{\sin\eta}{\eta}\right)^{1/2}.
\end{equation}
In evaluating Eq.~\eqref{eq:Gld int}, we note that $\eta\geq0$ since $\Delta t\geq0$.
The presence of the Heaviside function $\theta(\pi-\eta)$ can be understood as follows. For a
fixed $x\in\mt$, $\Gld$ is defined only for $x'\in\nn(x)$ [see Eq.~\eqref{eq:NN-x}] and has support
only when $\hat{\sigma}=0$. But for $\eta>\pi$ and $x'\in\nn(x)$, we have $\hat{\sigma}(x,x')<0$
and so $\Gld=0$. It is worth noting that $\Gld$ has compact support in $\eta$ due to the presence
of the two Heaviside $\theta$'s in Eq.~\eqref{eq:Gld}.  It is trivial to check that the original direct part,
Eq.~\eqref{eq:con-d}, is recovered by summing these modes as in
\begin{align}
  \label{eq:Green d}
  &
  \Gd(\coord{x},\coord{x'})=
\frac{1}{ r\, r'}
  \sum_{\ell=0}^{\infty}(2\ell+1)P_{\ell}(\cos\gamma)\Gld(r,r';\dt).
\end{align}
It is worth pointing out that $\Gd(x,x')$ is defined (via Eq.~\eqref{eq:con-d}) only for
$x'\in\nn(x)$. However, using the expression for $\Gld$ in (\ref{eq:Gld}), the sum on the right
hand side of (\ref{eq:Green d}) yields a quantity defined everywhere on $\mt$, which is identically zero for $x'$ outside $\nn(x)$.

Via Eq.~\eqref{eq:Gld} we have thus reduced the calculation of the direct multipolar modes $\Gld$
in Schwarzschild spacetime to the calculation of the distance $\eta$ and the van Vleck determinant
$\VVtd$ in $\Mtwo$. Unfortunately, closed form expressions for these quantities are not known and
so they must be calculated either: (i) numerically, as was done in Ref.~\cite{PhysRevD.92.104030},
by solving transport equations \cite{Ottewill:2009uj} along the unique geodesic that joins the two
points in $\Mtwo$; or (ii) using series approximations valid for small separations of the points
(see Appendix \ref{app:series}).\footnote{In the simpler cases of flat and Nariai spacetimes, the
equivalent of these quantities in their corresponding 2-D conformal spacetimes are known in closed
form and so the multipolar modes can also be obtained in closed form -- see Appendix A
in~\cite{casals2016global}.}

Summarizing, our method replaces the calculation of the retarded Green function in \eqref{eq:Green}
by that of the following ``non-direct" part of the retarded Green function,
\begin{align} \label{eq:Green nd}
&
\Gnd(\coord{x},\coord{x'})\equiv
\frac{1}{r\, r'}
\sum_{\ell=0}^{\infty}(2\ell+1)P_{\ell}(\cos\gamma)\left(\Glret(r,r';\dt)-\Gld(r,r';\dt)\right).
\end{align}
By construction, it is 
\begin{align}\label{eq:Gnd Had}
\Gnd(\coord{x},\coord{x'})=\Gret(\coord{x},\coord{x'})-\Gd(\coord{x},\coord{x'})=
\left\{\begin{array}{l l}
\Gret(\coord{x},\coord{x'})-U(\coord{x},\coord{x'})\delta(\sigma)\theta(\dt)=
V(\coord{x},\coord{x}')\theta(-\sigma)\theta(\dt), 
& \coord{x}'\in \mathcal{N}(x), \\
\displaystyle
\Gret(\coord{x},\coord{x'}),&\coord{x}'\notin \mathcal{N}(x).
\end{array}
\right.
\end{align}
Therefore,
$\Gnd(\coord{x},\coord{x'})$ is equal to $\Gret(\coord{x},\coord{x'})$ unless $\coord{x}'\in
\mathcal{N}(x)$ and $\sigma=0$. The advantage is that for $\sigma\neq 0$, Eq.~\eqref{eq:Green nd}
with the sum truncated at a given upper value $\ell=\ell_{\rm max}$ approximates the exact Green
function better than Eq.~\eqref{eq:Green} with the sum truncated at the same upper value. In
an abuse of language, we refer to $\Gnd$ as the ``regularized Green function".

\section{Demonstration of the method}
\label{sec:application}

We now present an explicit application of our method. For demonstration purposes we focus on the
case of a fixed point $\coord{x}$ with radial coordinate $r=6M$ connected to points $\coord{x'}$ by
a circular timelike geodesic. We emphasize, however, that the method works for any pair of points
in Schwarzschild spacetime.

We approximate (away from $\sigma=0$) the GF via Eq.~\eqref{eq:Green nd} with a truncated
sum. Specifically, we truncate the $\ell$-sum at $\ell_{\rm max}=100$ and, except where otherwise
specified, we include a smoothing factor $e^{-\ell^2/(2\ell_{cut}^2)}$ in the summands, with
$\ell_{\rm cut} = 20$. We evaluate $\Glret$ numerically using a surrogate model
\cite{Galley:Wardell:Gsur,GSur} generated from numerical data produced with the method described
in~\cite{PhysRevD.89.084021}. We also evaluate $\Gld$ numerically by solving transport equations,
as as was done in Ref.~\cite{PhysRevD.92.104030}. At early times, $0<t\lesssim 6 M$ in this
particular case, we encounter two problems:
\begin{enumerate}
  \item The time-domain numerical approach used to produce the data for the surrogate model for $\Glret$ works very well almost everywhere, but breaks down very near coincidence in the 2D spacetime (which corresponds to $r=r'$ and $\dt=0$, but not necessarily $\gamma=0$) where the numerical surrogate model (which essentially uses a 2D Gaussian approximation) used causes $\Glret$ to tend to $0$ rather than the true non-zero constant value at coincidence (as given in Eq.~\eqref{eq:local} below). This could be overcome by using a characteristic initial value formulation \cite{Mark:2017dnq,Marc:David,Conor:Barry:Adrian}   in place of the Gaussian approximation.
  \item There is significant cancellation between $\Glret$ and $\Gld$: both tend to the same non-zero, $\ell$-independent constant
    as $\dt \to 0$ [see Eqs.\eqref{eq:local} and \eqref{eq:local Gld} below]. 
   In order to achieve the expected result that $\Glret-\Gld = O(\dt)^6$ as $\dt \to 0$  [see Eq.\eqref{eq:Glret-Gld} below],
    we require increasing cancellation between $\Glret$ and $\Gld$ as $\dt$ decreases. At sufficiently small $\dt$ the numerical accuracy of $\Glret$ and $\Gld$ is insufficient to achieve this cancellation.
\end{enumerate}
We overcome both problems by using analytic, small-$\dt$ approximations to $\Glret$ and $\Gld$ in
place of numerics at early times. These approximations, while only valid for small-$\dt$, have the advantage that they accurately reproduce the near-coincidence behaviour and can be cancelled analytically without any concerns about numerical accuracy.  The details of these approximations are given in Appendices
\ref{app:series} and \ref{app:Bessel}. In the particular case studied here, we found the best
results when using the analytic appromations for $\Glret$ in the region $\dt < 6M$ and for $\Gld$
in the region $\dt < 3M$.

\subsection{Green function}

\begin{figure}[htb!]
\begin{center}
      \includegraphics[width=.48\textwidth]{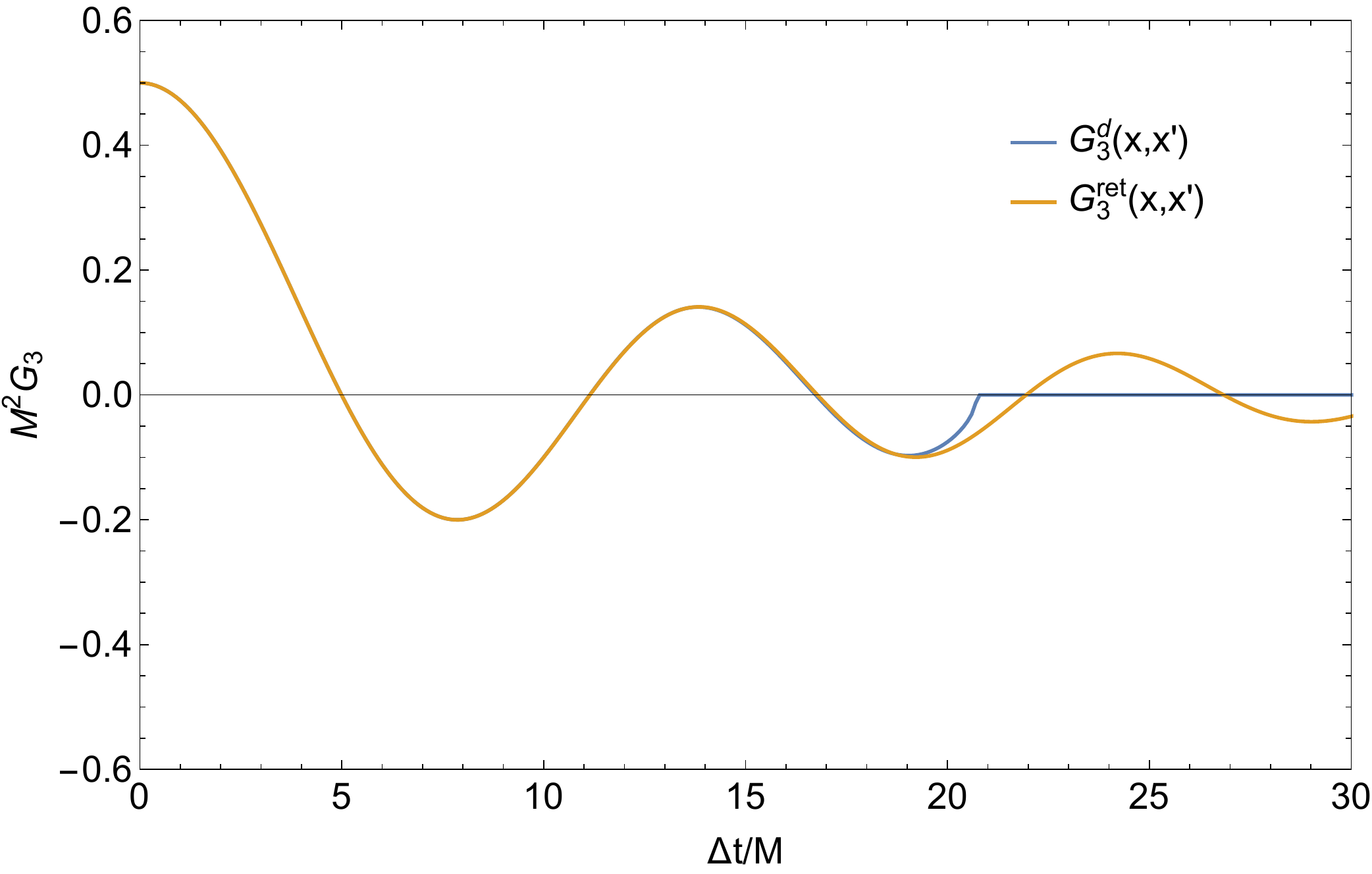}
      \includegraphics[width=.48\textwidth]{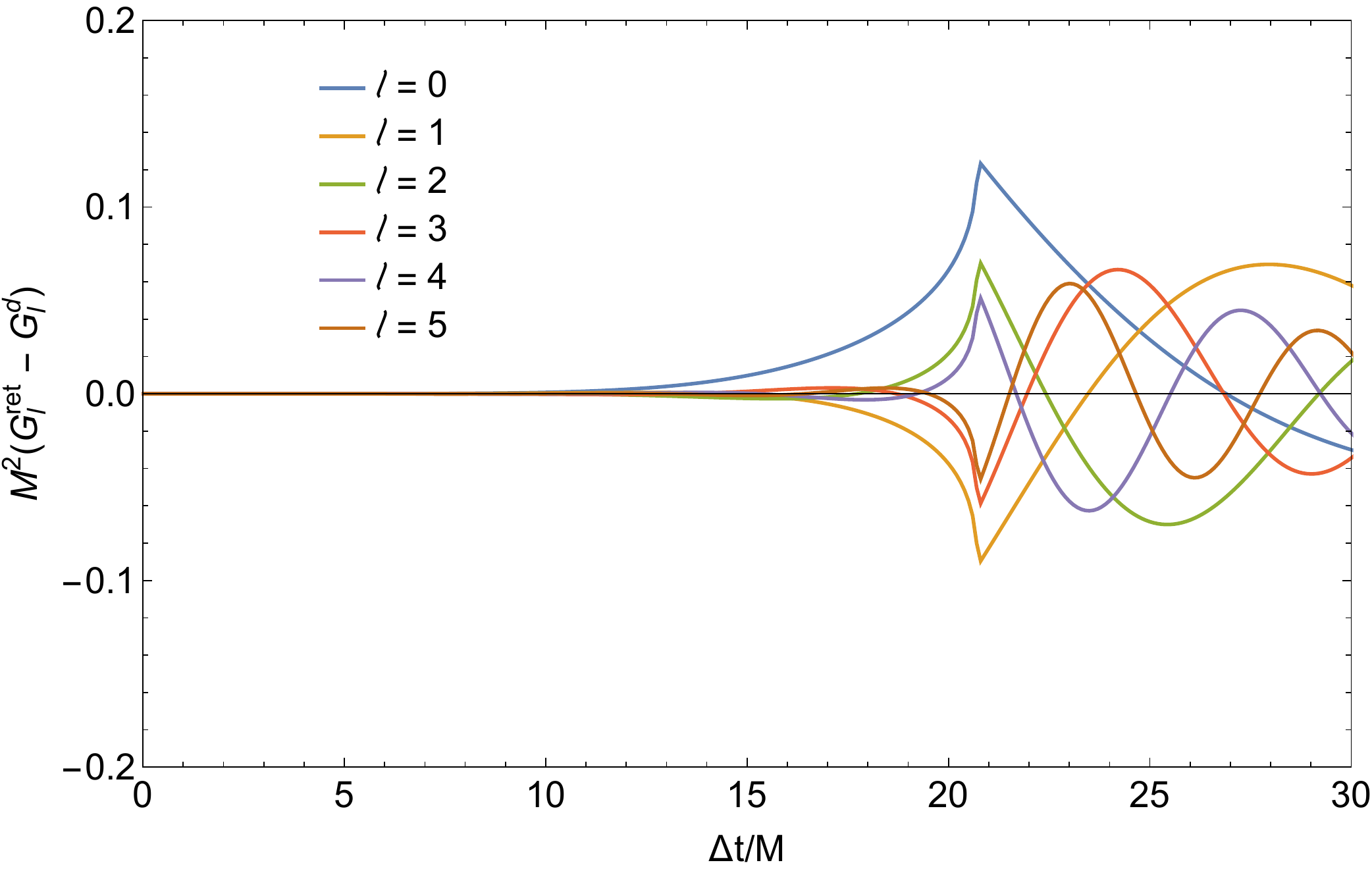}
\end{center}
\caption{
Green function $\ell$-modes for a scalar field on Schwarzschild spacetime as a function of time.
The point $\coord{x}$ is fixed at $r=6M$ and we vary the points $\coord{x'}$ along a circular
geodesic at $r=6M$.
Left: the red curve is $\Glret$ via the method in~\cite{PhysRevD.89.084021} and  the blue curve is $\Gld$ from Eq.~\eqref{eq:Gld} for $\ell=3$.
Right: $\Glret-\Gld$ for various $\ell$ values.
}
\label{fig:g3}
\end{figure}
We first study the behaviour of the regularized Green function. In Fig.~\ref{fig:g3} we plot the
modes $\Glret$, $\Gld$ and $\Glret-\Gld$. The left plot shows that $\Gld$ approximately agrees
with $\Glret$ throughout much of the region where $\Gld$ is non-zero. The right
plot shows the form -- including a breakdown in smoothness due to the compact support of $\Gld$ -- of the
last factor in the summand in Eq.~\eqref{eq:Green nd}.

\begin{figure}[htb!]
\begin{center}
  \includegraphics[width=.6\textwidth]{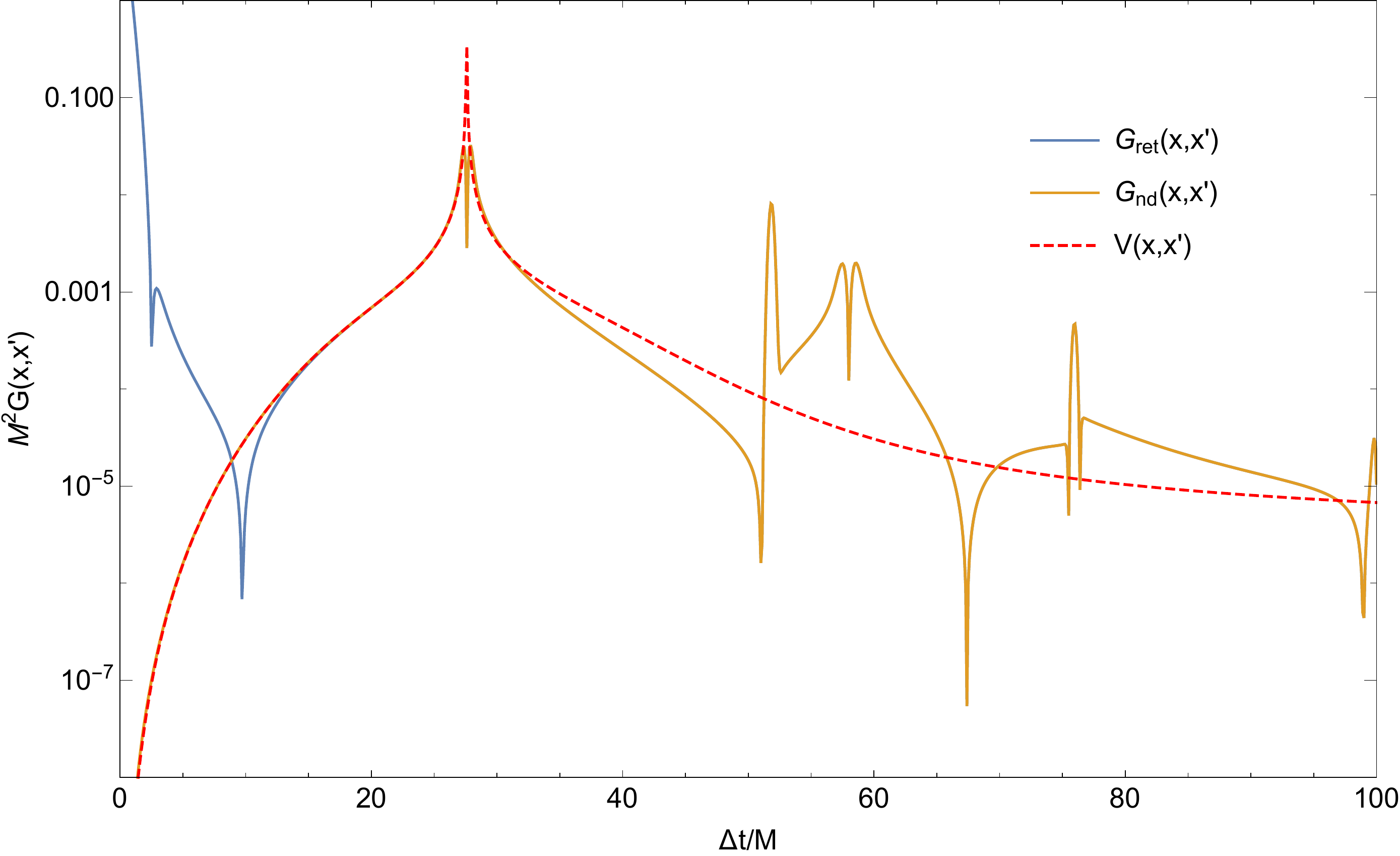}
\end{center}
\caption{
Retarded Green function $
\Gret(\coord{x},\coord{x}')$ for a scalar field on Schwarzschild spacetime as a function of time.
The point $\coord{x}$ is fixed at $r=6M$ and we vary the points $\coord{x'}$ along a circular
geodesic at $r=6M$. The solid blue line shows $\Gret$, Eq.~\eqref{eq:Green}, computed using a
truncated and smoothed multipolar mode-sum. The solid orange line shows
$\Gnd(\coord{x},\coord{x'})$, Eq.~\eqref{eq:Green nd}, also computed using a truncated and smoothed
multipolar mode-sum. In both cases smooth-sum parameters of $\ell_{\rm max}=100$ and $\ell_{\rm
cut} = 20$ were used. The dashed red line shows the Hadamard bitensor $V(\coord{x},\coord{x'})$
computed from a short distance Pad\'e-resummed Taylor series \cite{CDOWb}. }
\label{fig:GF}
\end{figure}
In Fig.~\ref{fig:GF} we plot the full $\Gret$ [Eq.~\eqref{eq:Green}], $\Gnd$ [Eq.~\eqref{eq:Green
nd}] and the Hadamard biscalar $V(x,x')$ [a Pad\'e resummation of Eq.~\eqref{eq:V power series} ---
see Ref.~\cite{CDOWb} for details]. The earliest time where the solid orange line --- which
corresponds to the calculation of $\Gnd$ --- spikes is where the first {\it non}-direct null
geodesic connects $\coord{x}$ to another point $\coord{x'}$ on the worldline. Since $\coord{x}$ and
this $\coord{x'}$ are connected by more than one causal geodesic (namely, the timelike worldline
and the first non-direct null geodesic) and this does not happen at earlier times, this time, which
is $\dt/M \approx 27.62$, marks the end of $\mathcal{N}(x)$ in this scenario. The other spikes in
the GF, which are located at $\dt/M \approx 51.84$, $58.05$, $75.96$ and $100.09$, correspond to
later non-direct null geodesics connecting $\coord{x}$ and other points $\coord{x'}$. The GF also
diverges at these points, but they are clearly outside $\mathcal{N}(x)$ and are of no relevance for
the specific purposes here. What is of relevance here is that the plot shows that, near
$\coord{x}$, Eq.~\eqref{eq:Green nd} performs much better than Eq.~\eqref{eq:Green}, as expected.

We end this subsection by making a comment about the behaviour at early times. Although not visible
in Fig.~\ref{fig:GF}, $\Gnd$ does not agree with $V(x,x')$ as well at early times ($\Delta t
\lesssim 6M$) as it does at later times. The reason for this slight disagreement can be understood
as follows. For $x'\in\mathcal{N}(x)$, the non-direct part $\Gnd$ is equal to
$V(x,x')\theta(-\sigma)\theta(\dt)$ [see \eqref{eq:Gnd Had}], not $V(x,x')$. One may expect this to
make no difference at early times (where $\sigma<0$ and $\dt >0$), but the fact is that we are not
computing the exact $\Gnd$, but an approximation to it, obtained by truncating the $\ell$-sum at
finite $\ell = \ell_{\rm cut}$ and by including a smoothing factor in Eq.~\eqref{eq:Green nd}. This
approximation ends up ``contaminating'' $\Gnd$ at early times. This can be seen explicitly when
carrying out a small-$\dt$ expansion in the case of the timelike circular geodesic at $r=6M$ that
concerns us. Making use of Eqs.~\eqref{eq:etasq} and \eqref{eq:Deltasq} and expanding Eqs.~\eqref{eq:Gld} and \eqref{eq:Glret Bessel} in a Taylor series
about $\dt=0$, we find (see Appendices \ref{app:series} and \ref{app:Bessel}):
\begin{align}
\label{eq:local}
  \Glret &= \tfrac{1}{2}-\tfrac{3 \ell^2+3 \ell+1}{1296}\dt^2 +\tfrac{18
   \ell^4+36 \ell^3+21 \ell^2+3 \ell-7}{6718464}\dt^4 -\tfrac{120 \ell^6+360 \ell^5+210
   \ell^4-180 \ell^3-591 \ell^2-441 \ell-67}{87071293440} \dt^6 + \cdots\\
   \label{eq:local Gld}
  \Gld &= \tfrac{1}{2}-\tfrac{3 \ell^2+3 \ell+1}{1296}\dt^2 +\tfrac{18
   \ell^4+36 \ell^3+21 \ell^2+3 \ell-7}{6718464}\dt^4 -\tfrac{840 \ell^6+2520
   \ell^5+1470 \ell^4-1260 \ell^3-4137 \ell^2-3087
   \ell-109}{609499054080} \dt^6 + \cdots
\end{align}
We thus find that these perfectly cancel through $\mathcal{O}(\dt^5)$, leaving a residual
contribution proportional to $\dt^6$,
\begin{equation}\label{eq:Glret-Gld}
  \Glret - \Gld = \tfrac{1}{1693052928} \dt^6+\cdots.
\end{equation}
Comparing against the small-$\dt$ expansion of $V(x,x')$ [using \eqref{eq:V power series}], which
in this case is given by
\begin{equation}\label{eq:V small dt}
  V(x,x') = -\tfrac{31}{13544423424} \dt^4 + \cdots,
\end{equation}
we find an apparent contradiction: $V(x,x') \sim \dt^4$ while $\Gnd \sim \dt^6$. The resolution of
this apparent contradiction is, as indicated above, that these are, in fact, not the exact
same quantity but differ by a factor of $\theta(-\sigma)\theta(\dt)$. By considering the mode 
decomposition of $\theta(-\sigma)\theta(\dt)$ times a small-$\Delta t$ expansion of $V(x,x')$, it is easily
verified that this difference accounts for the difference in two
orders of $\dt$ between the two expressions. 
Thus, the failure of  our approximation to $\Gnd$ to coincide with $V$
(and so with the GF) at early times can be attributed to the smooth sum approximating the step
function by a mollified version. Fortunately, this quirk has negligible effect on the results.

\subsection{Self-field}

A better approximation of the GF leads to a better approximation of the self-force and the
self-field. The self-force acting on a scalar charge $q$ moving on a worldline $z(\tau)$ on a
background spacetime (with $\tau$ proper time along the geodesic), is given by $f^{\mu}=q\nabla^{\mu}\Phi_R$,
where
\begin{equation}
  \label{eq:self-field}
  \Phi_R(\tau)\equiv \lim_{\epsilon\to 0^+} \int_{-\infty}^{\tau-\epsilon}d\tau' \Gret(z(\tau),z(\tau'))
\end{equation}
is the regularized self-field (the $\epsilon>0$ in the upper limit excludes the coincidence
$\coord{x'}=\coord{x}$, and so it excludes $\sigma=0$ inside the normal neighbourhood). 
One way of carrying out the integral is
 to match the
calculation of the GF via the multiple power series \eqref{eq:V power series} in the QL region to
that via the $\ell$-mode sum [either Eq.~\eqref{eq:Green nd} or Eq.~\eqref{eq:Green}] in the DP. This
requires a region of overlap and a choice of matching proper time $\tau_m$:
\begin{equation}
  \label{eq:self-field match}
  \Phi_R(\tau)=\int_{-\infty}^{\tau_m}d\tau' V(z(\tau),z(\tau'))+ 
  \frac{1}{r}
  \lim_{\epsilon\to 0^+}\int_{\tau_m}^{\tau-\epsilon}
  \frac{d\tau'}{r'} \sum_{\ell=0}^{\infty}(2\ell+1)P_{\ell}(\cos\gamma)\Glret(r,r';\dt),
\end{equation}
where $r'=r'(\tau')$, $t'=t'(\tau')$, $\gamma=\gamma(\tau')$; and similarly with $\Glret$ replaced
by $\Glret-\Gld$. The first integral in Eq.~\eqref{eq:self-field match} corresponds to the
DP contribution and the second integral to the QL contribution.

\begin{figure}[htb!]
\begin{center}
  \includegraphics[width=.6\textwidth]{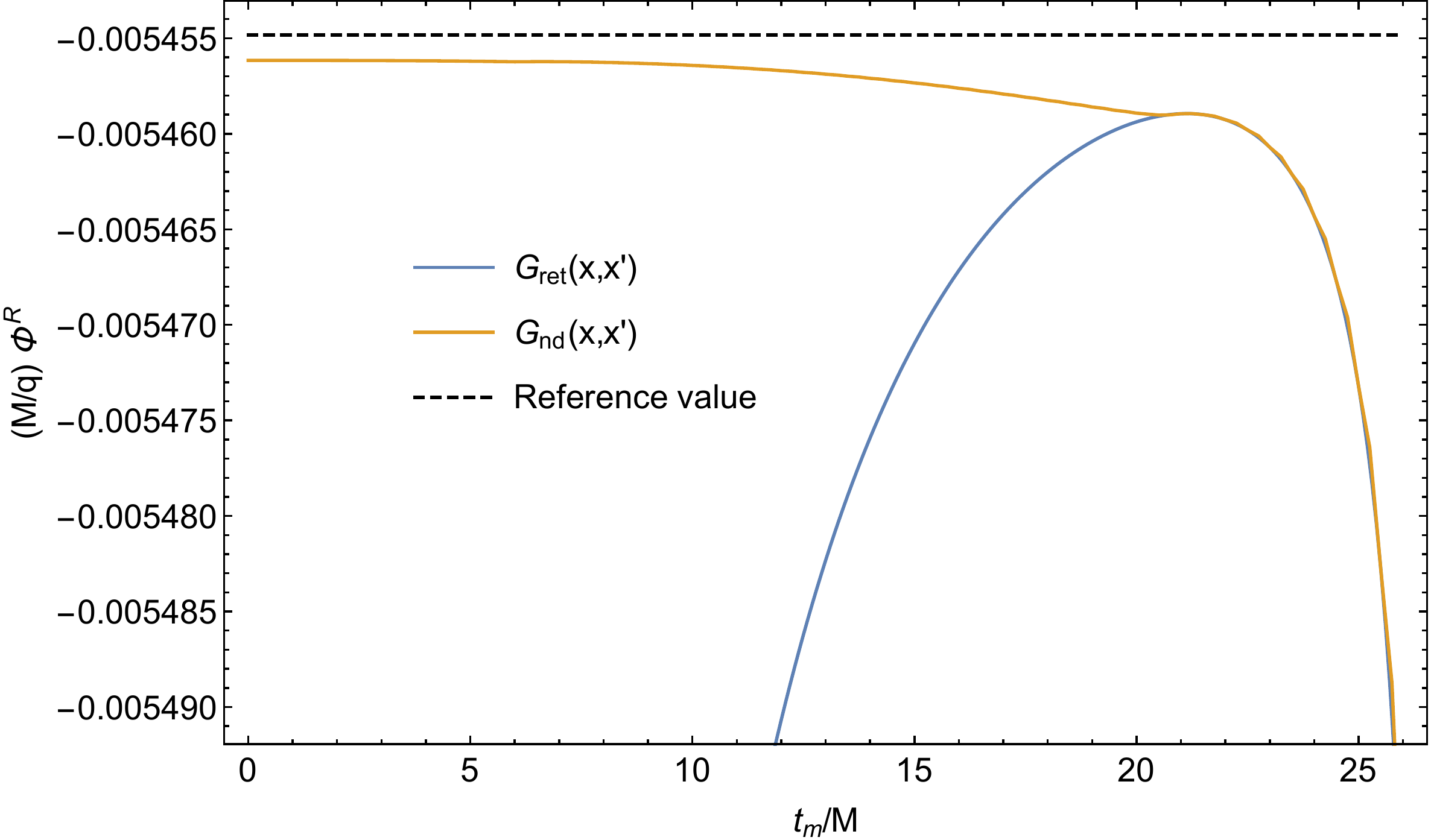}      
\end{center}
\caption{Plot of the regularized self-field $\Phi_R$
[calculated via Eq.~\eqref{eq:self-field match}] as a function of matching coordinate time $t_m$
for a scalar charge on a circular geodesic at $r=6M$. The solid blue line is obtained using
Eq.~\eqref{eq:self-field match} as it is (i.e., with $\Glret$ in the summand); the solid orange
line is obtained using Eq.~\eqref{eq:self-field match} with $\Glret$ replaced by
$\Glret-\Gld$ in the summand. In both cases the numerical integration was truncated at $t=200M$,
with the contribution for $t>200M$ accounted for by a late-time approximation which assumes the
branch-cut contribution to the GF dominates (see, e.g.,~\cite{CDOW13,PhysRevD.89.084021}). The dashed black line is a highly-accurate
reference value computed using the mode-sum regularization method \cite{Heffernan:2012su}. }
\label{fig:self-field}
\end{figure}

\begin{figure}[htb!]
\begin{center}
  \includegraphics[width=.52\textwidth]{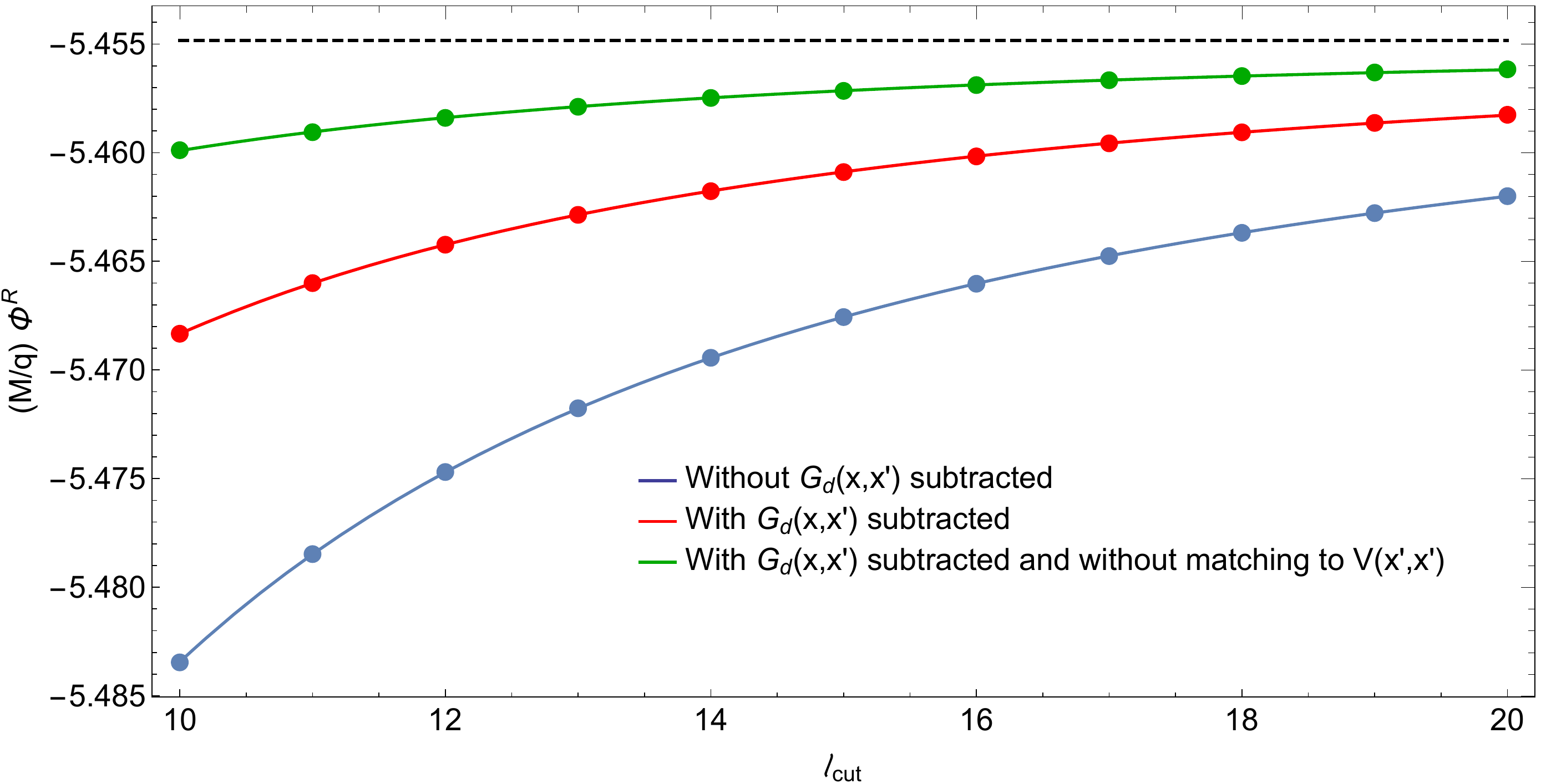}
\end{center}
\caption{The regularized self-field $\Phi_R$ for a scalar charge on a circular geodesic at $r=6M$
as a function of the parameter $\ell_{cut}$ in the smoothing factor in the $\ell$-sum (see
Sec.~\ref{sec:GF}). Blue curve: $\Phi_R$ is calculated via Eq.~\eqref{eq:self-field match} with
$\tau_m = 18M$. Red curve: $\Phi_R$ is calculated
via Eq.~\eqref{eq:self-field match} with $\tau_m=18M$ and with $\Glret$ replaced by $\Glret-\Gld$. Green curve: $\Phi_R$ is calculated
via Eq.~\eqref{eq:self-field match} with $\tau_m=0$ and with $\Glret$ replaced by $\Glret-\Gld$.}
\label{fig:self-field extrap}
\end{figure}

Fig.~\ref{fig:self-field} shows $\Phi_R$ for the case considered above of a scalar
charge on a circular geodesic at $r=6M$. We plot it as a function of the coordinate time $t_m$
corresponding to the matching proper time $\tau_m$, and compare the result to a
highly-accurate reference value computed using the mode-sum regularization method
\cite{Heffernan:2012su}. Similarly, in Fig.~\ref{fig:self-field extrap} we plot $\Phi_R$ as a
function of the parameter $\ell_{cut}$ in the smoothing factor in the $\ell$-sum (see
Sec.~\ref{sec:GF}). Both figures show that the calculation of $\Phi_R$ via Eq.~\eqref{eq:self-field
match} is much better when replacing $\Glret$ by $\Glret-\Gld$ than without the replacement. In
fact, from Fig.~\ref{fig:self-field} it is apparent that by using $\Glret-\Gld$ we have entirely
removed the need for matching to the power series approximation to $V(x,x')$.

\section{Discussion}
\label{sec:extension}

In this paper we have presented a proposal for facilitating the calculation of the retarded Green
function in Schwarzschild space-time. The proposal essentially consists of decomposing the Green
function into multipolar $\ell$-modes and significantly improving the convergence  of the sum by
subtracting the modes of the direct part in the Hadamard form. We have applied this method to the
case of a scalar charge and its self-field. We next discuss various interesting extensions of this
proposal and its applications.

First of all, we have applied this proposal to the calculation of the self-field, but the
calculation could be extended to the calculation of the self-{\it force}. This requires calculating
directly derivatives of the Green function, and so -- from Eq.~\eqref{eq:Gld} -- derivatives of the
van Vleck determinant and the world function in the 2-D conformal spacetime $\Mtwo$. A transport
equation prescription for obtaining derivatives of both the world function and the van Vleck
determinent are provided in~\cite{Ottewill:2009uj}.

Secondly, the proposal is readily generalizable from the zero-spin field considered here to
higher-spin fields, such as the electromagnetic field or the linear gravitational field. These
higher-spin fields in Schwarzschild space-time can be shown to satisfy a similar wave equation to
the (scalar) Klein-Gordon equation, merely with a change in the potential (but not in the
derivative terms). This implies that the retarded Green functions for these higher-spin wave
equations also admit the Hadamard form Eq.~\eqref{eq:hadamard}, with just a change in the biscalar
$V(x,x')$, but with the same world function $\sigma$ and biscalar $U(x,x')$. It also implies that
a conformal relationship similar to Eq.~\eqref{con-gf} between the Green functions in Schwarzschild and
conformal Schwarzschild space-times is satisfied for these higher-spin fields. Therefore, our proposed
Eq.~\eqref{eq:Green nd} also holds for these highers-spin cases, with the modes $\Glret$ and $\Gld$
satifying similar $(1+1)$-D wave equations as in the scalar case, but with different potentials.

Finally, in this paper we have focused on dealing with the singularity which the Green function
possesses at $\sigma=0$, i.e., due to the ``direct" null geodesic or at coincidence. However, as
mentioned, the Green function diverges when the points are connected by {\it any} null geodesic,
even if it is not the direct one. In~\cite{casals2016global}, the full (i.e., including leading and
subleading orders), global singularity structure of the Green function in Schwarzschild space-time
was provided (the leading order had been previously provided
in~\cite{Ori1short,Dolan:2011fh,harte2012caustics,Zenginoglu:2012xe}). When multipole-decomposing
the Green function, these non-direct singularities also arise as divergences in the multipolar
$\ell$-sum. Similarly to what we have done in Eq.~\eqref{eq:Green nd} for the direct singularity,
one could carry out $\ell$-mode decompositions of the non-direct singularities and subtract those
from the $\ell$-modes of the full Green function. One should note, however, that the non-direct
divergences alternate between Dirac-$\delta$ distributions (such as for the direct divergence) and
principal value distributions. For obtaining the $\ell$-modes, one would therefore have to perform
angular integrals of the principal value distribution instead of the Dirac-$\delta$ distribution.
It is expected that the resulting $\ell$-sum would then converge everywhere, thus greatly
faciliating further the calculation of the Green function. Regarding the subtracted non-direct
divergences, one could include them separately by calculating them using, e.g., the expressions
in~\cite{casals2016global}. Alternatively, if one is mainly interested in the calculation of the
self-field/force, which involves worldline-integrals of the Green function, one could subtract only
the part of the divergences which integrates to zero (i.e., with the coefficients of the diverging
functions, such as the principal value, evaluated at the times of the divergences), instead of the
full divergences (where the coefficients depend on time and so it would not integrate out to zero).


\section*{Acknowledgements}
M.~C.~acknowledges partial financial support by CNPq (Brazil), process number  310200/2017-2. This work makes use of the Black Hole Perturbation Toolkit \cite{BHPToolkit}.

\appendix

\section{Series expansions of geometric quantities in $\Mtwo$}
\label{app:series}

Equation \eqref{eq:local Gld} was obtained by substituting the expansion of the geometrical 
quantities $\epsilon\eta^2$ and $\Delta_{2d}^{1/2}$ through order $(\dx^a)^{6}$ into
Eq.~\eqref{eq:Gld}. For completeness we present these below; higher terms through
$(\dx^a)^{20}$ are provided electronically as supplemental material to this paper.
\begin{align}
\label{eq:etasq}
\epsilon\eta^2&=
-\frac{{\dt}^2 (r-2 M)}{r^3} 
+\frac{{\dr}^2}{r(r-2 M)}
+\frac{{\dr} {\dt}^2 (r-3 M)}{r^4}+\frac{{\dr}^3 (r-M)}{r^2 (r-2 M)^2}
-\frac{{\dt}^4 (r-2 M) (r-3 M)^2}{12 r^7}\nonumber\\&
-\frac{{\dr}^2 {\dt}^2 \left(5 r^2-28 M r+33 M^2\right)}{6r^5(r-2M)}
+\frac{{\dr}^4 \left(11 r^2-22 M r+15 M^2\right)}{12 r^3 (r-2 M)^3}
+\frac{{\dr} {\dt}^4 \left(2 r^3-20 M r^2+63 M^2 r-63 M^3\right)}{12 r^8}\nonumber\\&
+\frac{{\dr}^3 {\dt}^2 \left(4 r^3-31 M r^2+67 M^2 r-45 M^3\right)}{6 r^6  (r-2 M)^2}
-\frac{{\dr}^5 \left(10  r^3-30 M r^2+41 M^2 r-21 M^3\right)}{12 r^4 (r-2 M)^4}\nonumber\\&
-\frac{{\dt}^6 (r-3 M)^2 (r-2 M) \left(4  r^2-30 M r+45 M^2\right)}{360 r^{11}}
-\frac{{\dt}^4 {\dr}^2  \left(78 r^4-1072 M r^3+5087 M^2 r^2-10050 M^3 r+7065 M^4\right)}{360 r^9 (r-2 M)}\nonumber\\&
-\frac{{\dt}^2{\dr}^4  \left(64 r^4-606 M r^3+1781 M^2 r^2-2160 M^3 r+945 M^4\right)}{120 r^7 (r-2 M)^3}\nonumber\\&
+\frac{{\dr}^6 \left(274 r^4-1096 M r^3+2251 M^2 r^2-2310 M^3 r+945 M^4\right)}{360 r^5 (r-2 M)^5}+O\bigl((\dx^a)^7\bigr),
\end{align}
\begin{align}
\label{eq:Deltasq}
\Delta_{2d}^{1/2}&=
1 + 
\frac{{\dt}^2 (21 M^2 - 11 M r + r^2)}{6 r^4}
+ \frac{{\dr}^2 (15 M^2 - 11 M r + r^2)}{6 r^2 (r-2 M )^2}  
- \frac{ {\dr} {\dt}^2 (4 r^3 - 69 M r^2+ 290 M^2 r -342 M^3)}{
 24 (r-2 M ) r^5}  \nonumber\\
&
 + \frac{{\dr}^3 (4 r^3  - 69 M r^2 + 194 M^2 r -162 M^3 )}{24 r^3 (r-2 M)^3} 
 + \frac{ {\dt}^4 (r-2 M) (118 r^3-  - 2111 M r^2+ 9915 M^2 r-13680 M^3 )}{ 2160 r^8}  \nonumber\\
&
 +  \frac{{\dr}^2 {\dt}^2 (17 r^4 - 368 M r^3 + 2246 M^2 r^2  - 5064 M^3 r+3798 M^4 )}{108 r^6 (r-2 M)^2}  \nonumber\\
&
+ \frac{{\dr}^4 (38 r^4  - 897 M r^3 + 3887 M^2 r^2 - 6610 M^3 r+ 3960 M^4)}{240 r^4 (r-2 M)^4}  \nonumber\\
&
- \frac{{\dt}^4 {\dr}(1016 r^4 - 24575 M r^3  + 178346 M^2 r^2- 500670 M^3 r  + 477180 M^4 )}{8640 r^9}  \nonumber\\
&
+ \frac{{\dt}^2 {\dr}^3( 644 r^5- 16251 M r^4+ 125706 M^2 r^3 - 402955 M^3 r^2 + 571830 M^4 r-298890 M^5)}{4320 r^7 (r-2 M)^3}  \nonumber\\
&
+ \frac{{\dr}^5 ( 432 r^5  -  12973 M r^4 + 76588 M^2 r^3- 198220 M^3 r^2+ 240300 M^4 r-112140 M^5)}{2880 r^5 (r-2 M)^5}  \nonumber\\
&
+ \frac{{\dt}^6}{ 1814400 r^{12}} (29784 r^6  - 873125 M r^5   + 9135623 M^2 r^4 - 46052780 M^3 r^3   \nonumber\\
&\qquad\qquad\qquad\qquad\qquad\qquad + 121262184 M^4 r^2- 160751520 M^5 r + 84741660 M^6) \nonumber\\
&
+ \frac{{\dt}^4 {\dr}^2}{1814400 r^{10} (r-2 M)^2}  (304256 r^6 - 9729943 M r^5 + 106098469 M^2 r^4 -  541361342 M^3 r^3 \nonumber\\
&\qquad\qquad\qquad\qquad\qquad\qquad  + 1415223126 M^4 r^2 - 1839339360 M^5 r +  942984720 M^6 ) \nonumber\\
&
+ \frac{{\dt}^2 {\dr}^4}{1814400 r^8 (r-2 M)^4}   (258392 r^6  - 7409415 M r^5  + 68633901 M^2 r^4 -  281552428 M^3 r^3 \nonumber\\
&\qquad\qquad\qquad\qquad\qquad\qquad + 576945480 M^4 r^2 - 582918120 M^5 r+ 232438140 M^6
  ) \nonumber\\
    &
+ \frac{{\dr}^6}{
 201600 r^6 (r-2 M)^6} (28752 r^6  - 1049773 M r^5 + 7880183 M^2 r^4 -  27503250 M^3 r^3 \nonumber\\
&\qquad\qquad\qquad\qquad\qquad\qquad  + 50452170 M^4 r^2 - 47416440 M^5 r + 18078120 M^6  ) 
+O\bigl((\dx^a)^7\bigr).
\end{align}
In order to obtain Eq.~\eqref{eq:local Gld}, we evaluated
\eqref{eq:etasq} and \eqref{eq:Deltasq} for $\epsilon=-1$, $r=6M$ and $\Delta r=0$, inserted them
in Eq.~\eqref{eq:Gld} and re-expanded for small $\Delta t$.

\section{Bessel function expansion of $\Glret$}
\label{app:Bessel}
At early times, it is useful to have an analytic approximation to $\Glret$. This is easily obtained
using a small modification of the method described in Ref.~\cite{CDOWb}. In particular, we start
from the quantity $B(r,r')$ (which also depends on $\omega$ and $\ell$ defined in Eq.~(2.13) of Ref.~\cite{CDOWb}. Starting from the
expansion of $B(r,r')$ in powers of $r'-r$ and $\chi(r) \equiv [\omega^2 r^4 + r^2
f(r)(\ell+\frac12)^2]^{1/2}$ derived in that paper, we skip the sum over $\ell$ and proceed
directly to the inverse Fourier transform. This amounts to computing integrals over frequency
$\omega$ of the form
\begin{equation}\label{eq:Bessel}
  \int_0^\infty \chi^{-n-\frac12} \cos \left(\omega \dt \right) \, d \omega = - \frac{(-1)^n \sqrt{\pi} (i \dt)^n}{ r^{3n+2}f^{n/2} (2 \ell+1)^{n} \Gamma (n+\frac{1}{2})}I_{n}\left[\frac{\sqrt{f} (2\ell+1)}{2
   r} i \dt\right]
\end{equation}
where $n$ is a non-negative integer 
and $I_n(x)$ is the modified Bessel function of the first kind. The result
is an expression for $B(r,r')$  (and thus $\Glret$) as an infinite series of Bessel functions.
Including a given number, $n$, of terms in the series yields a result which is accurate through
$\mathcal{O}(\dt^{2n})$ (note, however, that it is better to keep the Bessel function form as that
gives a more accurate result over a larger range of values for $\dt$). Explicitly, in the case
$r=r'$ the first few terms are
\begin{align}\label{eq:Glret Bessel}
  \Glret &= 
    \frac{1}{2}I_0
    -\frac{i \dt \, I_1 }{8 (2 \ell+1) r^2} \sqrt{f} (r-8 M)
    +\frac{\dt^2 \, I_2}{192 (2 \ell+1)^2 r^5}\Big[32 (30 \ell^2+30 \ell-79) M^2 r-2 (320 \ell^2+320 \ell-219) M r^2 \nonumber \\ & \qquad
       +3 (32 \ell^2+32 \ell-1) r^3+3456 M^3\Big]
    -\frac{i \dt^3 \, I_3}{3840 \sqrt{f} (2 \ell+1)^3 r^8} \Big[11520 (54 \ell^2+54 \ell-281) M^4 r+160 (72 \ell^4+144
   \ell^3 \nonumber \\ & \qquad
   -5044 \ell^2-5116 \ell+10191) M^3 r^2-4 (3360 \ell^4+6720 \ell^3-91280 \ell^2-94640
   \ell+86013) M^2 r^3+4 (1280 \ell^4\nonumber \\ & \qquad
     +2560 \ell^3-16464 \ell^2-17744 \ell+6275) M r^4-5
   (128 \ell^4+256 \ell^3-736 \ell^2-864 \ell+1) r^5+2304000 M^5\Big] + \cdots,
\end{align}
where the argument of all the Bessel functions is the same as
that of $I_n$ in Eq.~\eqref{eq:Bessel}. In this work, we make use of an expansion through $n=20$;
the higher terms are provided electronically as supplemental material to this paper. In order to obtain Eq.~\eqref{eq:local}, we evaluated \eqref{eq:Glret
Bessel} for $r=6M$ and expanded for small $\Delta t$.


\bibliographystyle{apsrev}


\end{document}